# Nanogrid Power Management Based on Fuzzy Logic Controller

*Walid Issa[1*], Rashid Albadwawi[2], Tedjani MESBAHI[3], Mohammad Abusara[2]*

[1]*Sheffield Halam University, Sheffield, UK*
[2]*University of Exeter, Cornwall, UK*
[3]*INSA Strasbourg, Strasbourg, France*
*\*walid.issa@shu.ac.uk*

## Abstract

The transport industry is undergoing unprecedented technological change. In the automotive sector, the voices of progress are, among other things, linked to the partial or total electrification of vehicles. Thanks to bi-directional charging technology, electric vehicles (EV) could significantly improve their homeowners' energy balance. For islanded household buildings where a supplementary turbine generator and photovoltaic (PV) generator are the main supplies, it is crucial to minimize the running costs and maximize the use of the PV power. In this paper, a supervisory controller based on fuzzy logic is proposed to utilize the battery in the EV, while parking, as an interactive element to replace the auxiliary turbine and to assure that the battery power and energy do not exceed their design limits and maintaining a stable power flow. The nanogrid considered in this paper consists of a PV, EV battery, load and a turbine supplementary unit. The fuzzy logic controller alters the AC bus frequency, which is used by the local controllers of the parallel units to curtail the power generated by the PV or to supplement the power from the turbine unit. The proposed FLC performance is verified by Matlab simulation.

## 1. Introduction

In the first half of 2020, renewable energies accounted for more than 50% of total electricity production in Germany for the first time. This is a new record and paves the way for the use of intelligent renewable energy solutions. In this context, the concept of the microgrids and nanogrids have emerged in response to the increased penetration of renewable energy systems (RES). In a microgrid, distributed generation (DG) units, energy storage systems (ESS), and loads are aggregated as one unit connected to the grid or isolated. Due to their controllability, microgrids will become the building blocks of future smart grids. Nanogrids are considered as a smaller scale of microgrids typically serving a single customer or facility. Low voltage grids with high penetration of PV has an impact on the grid health related to voltage magnitude variations along the feeders. These variations are particularly evident during periods of high generation and low load conditions [1]. Such events may occur daily in residential areas with high concentration of roof-top PV. Several options for voltage rise mitigation have been investigated and the most important include: grid reinforcement [2], coordinated active power curtailment [3], reactive power control strategies [4], and stationary storage [5].

Programs that aim to accelerate the development or increase the market share of electric vehicles can be found in several countries. The governments' motivations and the market investments toward EVs will lead eventually to a wide spread of EV using the grid for charging. If the charging of EVs is supported by more intelligent charging schemes that take into account the actual production of RES, both the environmental value and the electric energy sector value may grow significantly [6]. In household building nanogrids, Fig. 1, where the micro gas turbine (μGT) is essentially used and the PV generation supports supplying higher loads, the μGT needs to run always regardless of the demand and available PV power. Furthermore, the excess PV power can't be stored for a later use. With more EVs will be available, the use of EVs as a storage solution is the motivation of this research. Moreover, by this topology the use of μGT will be ceased in the case of EV is plugged in with a healthy state. According to Dallinger et al. [7] and Weiller et al. [8], the time interval between 10:00 AM and 15:00 PM shows a lower driving activity compared to the early morning interval 7:00 AM to 9:00 AM, which may correspond to a higher probability of EV connected to the household network. The EV batteries could discharge the stored electricity to the nanogrid on demand, which is collectively termed as the Vehicle-to-Building (V2B) concept. Thus, the EV storage could operate as a controllable load or distributed generator [9]. Vehicle-to-building and vehicle-to-grid service opportunities are envisioned [10]. Charging strategies such as individual peak shaving and droop-based voltage support [11] can be considered as support mechanisms for buildings and grids.

In the literature, the focus of EV power coordination in buildings requires the future knowledge of the EV state and usage, building power consumption and/or electricity cost. In addition, real-time control of the EV power has mostly resorted to a deterministic strategy where the PV output, EV power, and loads are assumed to be known in advance [12]. However, EV coordination strategies in larger buildings with limited prior knowledge and communication, or grid stabilizing strategies [11] are not extensively discussed. In [13], a control combines fuzzy control with gain scheduling techniques to accomplish both power sharing and energy management. In [14], an economic dispatch problem for total operation cost minimization in DC microgrids is formulated and solved with a heuristic method. However, these approaches address grid-connected microgrids. In [15], centralized energy management system (EMS) for isolated nanoogrids are designed by decomposing the unit commitment and optimal power flow in order to avoid a mixed-integer non-linear formulation. However, most of these proposals are not tested experimentally. A frequency bus-signalling technique of ESS is also used in [16] to manage an islanded AC microgrid with a PV, ESS and load. It is achieved by mapping AC bus frequency with estimated SOC.

In this paper, an FLC is proposed to manage an islanded household building AC nanogrid with a PV, EV and an µGT units. It prevents the EV battery SOC and charging/ discharging power from exceeding their limits regardless of the variation in the load and intermittent power generated by the RES. It is worth mentioning here that the focus of the study is only on the islanded mode of the nanogrid. The main contributions can be emphasized when the nanogrid is islanded. In grid-connected mode, each unit behaves differently without having the same concerns which this paper addresses. By varying the AC bus frequency that is used by local droop controllers; the FLC, located in the EV, is implemented without the need for any communication links between the nanogrid units. The FLC decides whether to curtail the power generated by the PV or to supplement the power from the auxiliary unit. Real time simulation validation of the proposed controller under different scenarios of operation and a comparison with a traditional controller. Moreover, the proposed EMS is experimentally tested in a laboratory scale nanogrid. Usually, the diesel generators or gas turbines are used as the main sources which dictates the AC bus. However, the auxiliary unit (μGT) here is floating and hence it provides power via the FLC command whenever needed (low power from RES or/and low SOC) which reduces the total running cost.

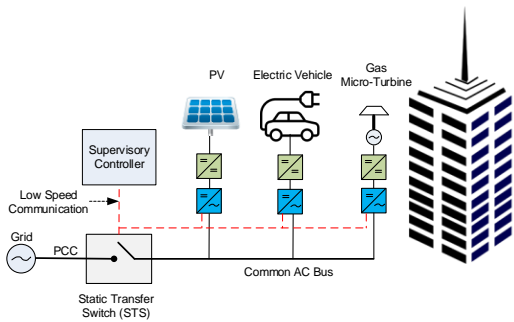

Fig. 1. General nanogrid structure under study

The nanogrid presented in Fig. 2 is selected as case study to validate the proposed EMS. This MG is composed of PV-based RES unit, an EV battery and µGT unit, that supply a common load bus. The considered study is when the is plugged into the building. Voltage-source converters are used for interfacing the DG and the EV battery with the nanogrid [17].

## 2. Case Study

The proposal assumes that the EV is connected to the building network. The proposed stand-alone AC building-nanogrid control topology is shown in Fig. 2 and it operates as follows.

1) PV unit is interfaced by a uni-directional DC/DC converter and a DC/AC inverter. The converter controls the PV output voltage to achieve maximum power point tracking (MPPT) while the inverter regulates the DC link voltage. The PV output power is curtailed if the battery is fully charged and the available PV power is higher than that required by the load.

2) EV battery unit is interfaced by a bi-directional DC/DC converter and a DC/AC inverter. The converter regulates the DC link voltage. The inverter is the master unit that maintains and controls the AC bus frequency and voltage of the nanogrid. It alters the bus frequency according to FLC command. The EV unit forms the AC bus by controlling the local AC voltage and frequency. The EV battery absorbs surplus power from the PV unit if it exceeds the load. In the same way, the EV battery supply required power when there is a shortage from the PV unit that can't meet the load requirement.

3) Auxiliary unit (micro gas turbine in this case) is interfaced by a uni-directional AC/DC converter and a DC/AC inverter. The converter regulates the DC link voltage while the inverter controls the output power according to the AC bus frequency altered based on FLC command. The main role of the auxiliary unit is to support the EV battery unit during low battery SOC and/or low PV generation scenarios that can't meet the load requirement.

Droop control [18], [19] is used in all the three DC/AC inverters as a frequency responsive technique. The details of power management based on droop control and bus frequency signaling techniques will be discussed in the next section. It is important to note that the supervisory control just requires to communicate with the EV battery to manage the nanogrid powers.

## 3. Fuzzy Logic Controller -based EMS

Fuzzy logic is designed with several IF and THEN rules based on human knowledge and experience. It could be appropriate option with complex systems like nanogrid with different types of inputs, variables and disturbances in particular if they are connected or supplied through RES. The proposed FLC is responsible for varying the bus frequency and is shown in Fig. 3. It consists of two subsystems. The top subsystem is responsible for preventing the battery from overcharging and its charging power from exceeding its limit. The inputs for this subsystem are $\Delta SOC_1$ and $\Delta P_{charge}$ which are given by (1) and (2), respectively.

$$\Delta SOC_1 = \frac{SOC^*_{max} - SOC}{SOC^*_{max} - SOC^*_{min}} \qquad (1)$$

$$\Delta P_{charge} = \frac{P^*_{charge\_max} - P_{charge}}{P^*_{charge\_max}} \qquad (2)$$

where $SOC$ is the current state of charge and $SOC^*_{max}$ is its maximum value. $P_{charge}$ is charging power and $P^*_{charge\_max}$ is its maximum charging power value. The output is a positive shift in the frequency $\Delta\omega_+$. As this controller is implemented in the EV battery, the bus frequency will deviate to a new frequency. In response to that the PV power can be curtailed. On the other hand, the bottom FLC subsystem is responsible for preventing the battery from over-discharging and the battery discharging power from exceeding its limit. The inputs for this subsystem are $\Delta SOC_2$ and $\Delta P_{discharge}$ which are given by (3) and (4), respectively.

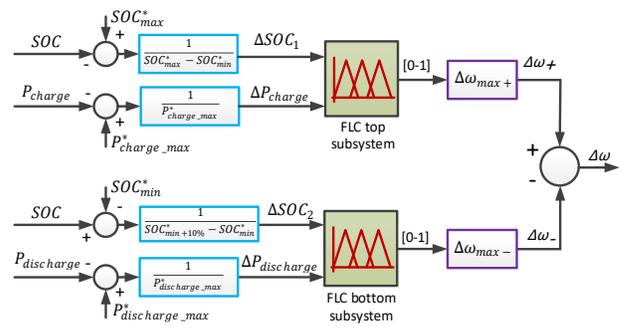

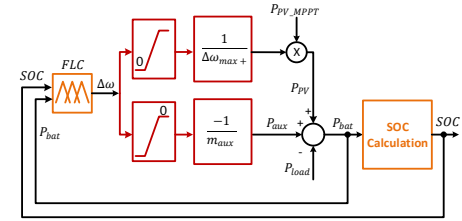

Fig. 3. Proposed fuzzy logic controller.

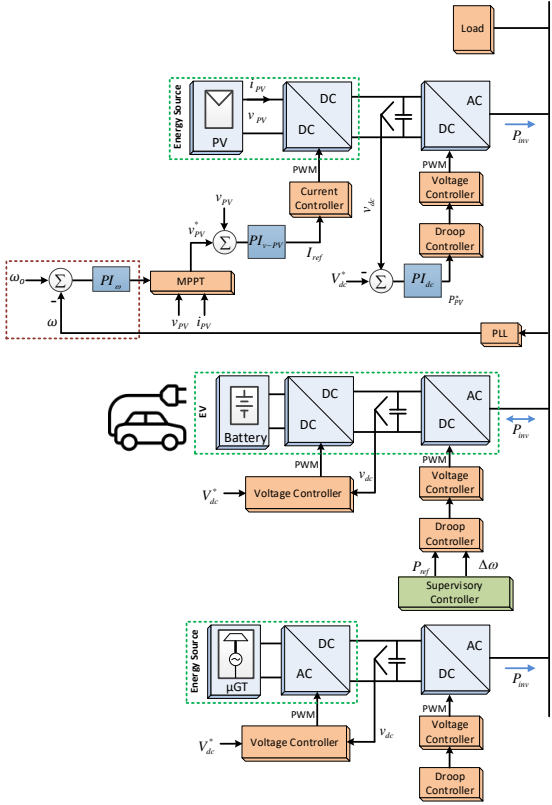

Fig. 2. Proposed stand-alone AC nanogrid control topology

$$\Delta SOC_2 = \frac{SOC - SOC^*_{min}}{SOC^*_{min+10\%} - SOC^*_{min}} \quad (3)$$

$$\Delta P_{discharge} = \frac{P^*_{discharge\_max} - P_{discharge}}{P^*_{discharge\_max}} \quad (4)$$

where $SOC^*_{min}$ is the $SOC$ minimum value and $SOC^*_{mid+10\%}$ is the $SOC$ minimum value plus 10%. $P_{discharge}$ is discharging power and $P^*_{discharge\_max}$ is its maximum discharging power value. The output is a negative shift in the frequency $\Delta\omega_-$ which deviates the bus frequency causing the auxiliary unit to supplement power.

Fig. 4 Simplified simulation model.

## 4. Real-Time Simulation Results

A simplified model of the AC nanogrid consists of a PV, battery and auxiliary units along with the proposed controllers has been developed and built in Matlab/Simulink and Fuzzy Logic tool boxes using RT-LAB (real-time simulator).

The first scenario represents a battery having high SOC with initial value approaching the maximum limit of 95%. Fig. 5(a) shows the power output of the PV, battery and auxiliary units along with the load power. The expectation is that the PV power should be used to supply the load and any excess power will be curtailed. Initially, there is no power generated by the PV since the solar radiation is almost zero during the first 30min. The battery is completely supplying the load (starting from 200W) and the auxiliary unit is not supplying any power as the battery SOC is high. After the PV starts generating more power, the contribution from the battery is reduced. At t=1h, the PV generation is almost following the load's profile and the extra power is curtailed. Most of the time, the used PV is a little bit higher than the load as shown in Fig. 5(a) and the battery is not really used much as per the FLC command since the priority is given for full utilization of PV power and the SOC is high. However, whenever there is a need for extra power to meet the load, the battery is supplying that extra power and this can be easily observed from t=9.5h onwards. The auxiliary unit is not used at all throughout the simulation since the PV and SOC of the battery are sufficient to cope with the load. The charging/discharging power is maintained within its limit. Fig. 5(b) shows that the SOC remains almost constant and it is prevented from exceeding its maximum limit. Fig. 5(c) shows the frequency curve where the frequency is maintained within its limits as well irrespective of the changes in the load or the PV generation.

The second scenario describes the case when the battery has a low SOC with initial value equals to the minimum limit value (40%). However, this time the load profile is multiplied by a factor of 4 to have higher values for simulation with similar trend and the rest remains the same. This helps in having a wide range of load to check the performance of the FLC. In this scenario, most of the time, the available PV power is lower than the load profile which means there is more need for support from the battery and auxiliary unit. Fig. 6(a) shows the power output along with the load power. The FLC has the decision to run the auxiliary unit to provide power to charge the battery and to supply the load if the PV power is low. After the first half an hour, the PV starts generating power. Since the SOC value is low, the battery is straight away in charging mode using the auxiliary unit to avoid possible decline of the SOC value to a value less than the SOC minimum allowable limit (40%). The auxiliary unit is floating throughout the simulation period and providing the required power as per the FLC command. It is obvious that the maximum charging/discharging power of the battery is well preserved within the maximum allowable limit (1000W) throughout the full period of the simulation. Fig. 6(b) shows the SOC curve which reflects good performance towards increasing the SOC regardless of the generation and demand variations. The EV battery is mostly in charging mode. Fig. 6(c) shows that frequency is maintained within its limits as well.

TABLE 1
SIMULATED SYSTEM PARAMETERS

| Parameter | Symbol | Value |
|---|---|---|
| PV power rating | $P_{pv}$ | 2230W |
| Auxiliary power rating | $P_{aux}$ | 1000W |
| Battery capacity | $C_{bat,}$ | 100Ah |
| Battery voltage | $V_{bat,}$ | 120V |
| Maximum state of charge | $SOC^*_{max}$ | 95% |
| Minimum state of charge plus 10% | $SOC^*_{min+10\%}$ | 50% |
| Minimum state of charge | $SOC^*_{min}$ | 40% |
| Maximum charging power | $P^*_{charge\_max}$ | 1000W |
| Maximum discharging power | $P^*_{discharge\_max}$ | 1000W |
| Nominal bus frequency | $\omega_o$ | 314.16rad/s |
| Active power droop coefficients | $m_{pv}, m_{aux}$ | 0.75e-4 rad/s/W |
| Reactive power droop coefficients | $n$ | 0.75e-4 V/Var |

## 5. Conclusion

An energy management system based on fuzzy logic control has been proposed for smart building application where EV can be used at times when the vehicle is parking. The proposed controller combines the FLC and bus signaling techniques to control the power flow between different energy sources. By varying the AC bus frequency, within the standards allowance, and making use of local droop controllers, the controller is implemented without any communication links between the nanogrid units. The RES unit can properly react to curtail its power when needed and the auxiliary unit is left to float on the AC bus so it reacts instantaneously to frequency variation to supply power. The results showed that the proposed FLC is capable to satisfy the system requirements within the defined constrains. It maintains the SOC and charging/discharging power of the battery within their limits irrespective of the change in RES/load. The performance has been validated by real time simulation and experimentally. The performance of the FLC is superior to the performance when compared to a traditional droop control method (i.e. proportional controller in this case) in achieving the required goals. The proportional controller was not always able to maintain the SOC and charging/discharging power within their design limits.

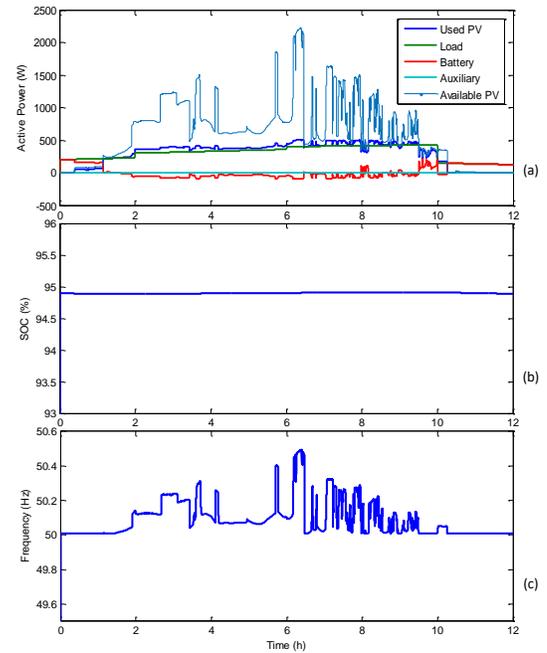

Fig. 5. Output response for 94.9% SOC case: (a) power (b) frequency (c) SOC.

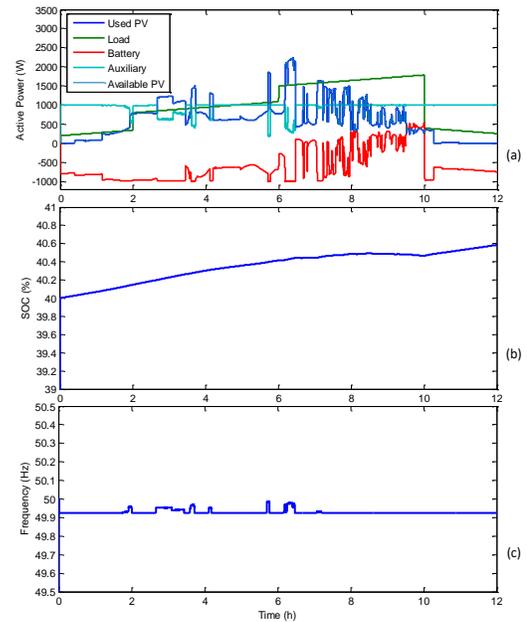

Fig. 6. Output responses for 40% SOC case and high load: (a) power (b) frequency (c) SOC.